\documentclass[letter,traditabstract,longauth]{aa}
\usepackage{graphicx}
\usepackage{txfonts}

\begin{document}
   \title{HerMES: The SPIRE confusion limit\thanks{{\it Herschel} is an ESA space observatory with  
science instruments provided by by European-led Principal Investigator consortia and with important  
participation from NASA.}}

\author{H.\,T.~Nguyen\inst{1,2}
\and B.~Schulz\inst{2,3}
\and L.~Levenson\inst{2}
\and A.~Amblard\inst{4}
\and V.~Arumugam\inst{5}
\and H.~Aussel\inst{6}
\and T.~Babbedge\inst{7}
\and A.~Blain\inst{2}
\and J.~Bock\inst{1,2}
\and A.~Boselli\inst{8}
\and V.~Buat\inst{8}
\and N.~Castro-Rodriguez\inst{9}
\and A.~Cava\inst{9}
\and P.~Chanial\inst{7}
\and E.~Chapin\inst{10}
\and D.\,L.~Clements\inst{7}
\and A.~Conley\inst{11}
\and L.~Conversi\inst{12}
\and A.~Cooray\inst{4,2}
\and C.\,D.~Dowell\inst{1,2}
\and E.~Dwek\inst{13}
\and S.~Eales\inst{14}
\and D.~Elbaz\inst{6}
\and M.~Fox\inst{7}
\and A.~Franceschini\inst{15}
\and W.~Gear\inst{14}
\and J.~Glenn\inst{11}
\and M.~Griffin\inst{14}
\and M.~Halpern\inst{10}
\and E.~Hatziminaoglou\inst{16}
\and E.~Ibar\inst{17}
\and K.~Isaak\inst{14}
\and R.\,J.~Ivison\inst{17,5}
\and G.~Lagache\inst{18}
\and N.~Lu\inst{2,3}
\and S.~Madden\inst{6}
\and B.~Maffei\inst{19}
\and G.~Mainetti\inst{15}
\and L.~Marchetti\inst{15}
\and G.~Marsden\inst{10}
\and J.~Marshall\inst{2,1}
\and B.~O'Halloran\inst{7}
\and S.\,J.~Oliver\inst{20}
\and A.~Omont\inst{21}
\and M.\,J.~Page\inst{22}
\and P.~Panuzzo\inst{6}
\and A.~Papageorgiou\inst{14}
\and C.\,P.~Pearson\inst{23,24}
\and I.~Perez Fournon\inst{9}
\and M.~Pohlen\inst{14}
\and N.~Rangwala\inst{11}
\and D.~Rigopoulou\inst{23,25}
\and D.~Rizzo\inst{7}
\and I.\,G.~Roseboom\inst{20}
\and M.~Rowan-Robinson\inst{7}
\and Douglas~Scott\inst{10}
\and N.~Seymour\inst{22}
\and D.\,L.~Shupe\inst{2,3}
\and A.\,J.~Smith\inst{20}
\and J.\,A.~Stevens\inst{26}
\and M.~Symeonidis\inst{22}
\and M.~Trichas\inst{7}
\and K.\,E.~Tugwell\inst{22}
\and M.~Vaccari\inst{15}
\and I.~Valtchanov\inst{12}
\and L.~Vigroux\inst{21}
\and L.~Wang\inst{20}
\and R.~Ward\inst{20}
\and D.~Wiebe\inst{10}
\and G.~Wright\inst{17}
\and C.\,K.~Xu\inst{2,3}
\and M.~Zemcov\inst{2}}

\institute{Jet Propulsion Laboratory, 4800 Oak Grove Drive, Pasadena, CA 91109, USA\\
 \email{htnguyen@jpl.nasa.gov}
\and California Institute of Technology, 1200 E.\ California Blvd., Pasadena, CA 91125, USA
\and Infrared Processing and Analysis Center, MS 100-22, California Institute of Technology, JPL, Pasadena, CA 91125, USA
\and Dept.\ of Physics \& Astronomy, University of California, Irvine, CA 92697, USA
\and Institute for Astronomy, University of Edinburgh, Royal Observatory, Blackford Hill, Edinburgh EH9 3HJ, UK
\and Laboratoire AIM-Paris-Saclay, CEA/DSM/Irfu - CNRS - Universit\'e Paris Diderot, CE-Saclay, pt courrier 131, F-91191 Gif-sur-Yvette, France
\and Astrophysics Group, Imperial College London, Blackett Laboratory, Prince Consort Road, London SW7 2AZ, UK
\and Laboratoire d'Astrophysique de Marseille, OAMP, Universit\'e Aix-marseille, CNRS, 38 rue Fr\'ed\'eric Joliot-Curie, 13388 Marseille cedex 13, France
\and Institute de Astrofisica de Canarias, C/ Via Lactea s/n, E-38200 La Laguna, Spain
\and Department of Physics \& Astronomy, University of British Columbia, 6224 Agricultural Road, Vancouver, BC V6T~1Z1, Canada
\and Dept.\ of Astrophysical and Planetary Sciences, CASA 389-UCB, University of Colorado, Boulder, CO 80309, USA
\and Herschel Science Centre, European Space Astronomy Centre, Villanueva de la Ca\~nada, 28691 Madrid, Spain
\and Observational  Cosmology Lab, Code 665, NASA Goddard Space Flight  Center, Greenbelt, MD 20771, USA
\and Cardiff School of Physics and Astronomy, Cardiff University, Queens Buildings, The Parade, Cardiff CF24 3AA, UK
\and Dipartimento di Astronomia, Universit\`{a} di Padova, vicolo Osservatorio, 3, 35122 Padova, Italy
\and ESO, Karl-Schwarzschild-Str.\ 2, 85748 Garching bei M\"unchen, Germany
\and UK Astronomy Technology Centre, Royal Observatory, Blackford Hill, Edinburgh EH9 3HJ, UK
\and Institut d'Astrophysique Spatiale (IAS), b\^atiment 121, Universit\'e Paris-Sud 11 and CNRS (UMR 8617), 91405 Orsay, France
\and School of Physics and Astronomy, The University of Manchester, Alan Turing Building, Oxford Road, Manchester M13 9PL, UK
\and Astronomy Centre, Dept.\ of Physics \& Astronomy, University of Sussex, Brighton BN1 9QH, UK
\and Institut d'Astrophysique de Paris, UMR 7095, CNRS, UPMC Univ.\ Paris 06, 98bis boulevard Arago, F-75014 Paris, France
\and Mullard Space Science Laboratory, University College London, Holmbury St.\ Mary, Dorking, Surrey RH5 6NT, UK
\and Space Science \& Technology Department, Rutherford Appleton Laboratory, Chilton, Didcot, Oxfordshire OX11 0QX, UK
\and Institute for Space Imaging Science, University of Lethbridge, Lethbridge, Alberta, T1K 3M4, Canada
\and Astrophysics, Oxford University, Keble Road, Oxford OX1 3RH, UK
\and Centre for Astrophysics Research, University of Hertfordshire, College Lane, Hatfield, Hertfordshire AL10 9AB, UK}

   \date{Received Apr 6, 2010; accepted Apr 20, 2010}

 
\abstract
   {We report on the sensitivity of SPIRE photometers
 on the {\it Herschel} Space Observatory.   Specifically, we measure the confusion noise from observations taken during the Science Demonstration Phase of the {\it Herschel} Multi-tiered Extragalactic Survey.  Confusion noise is defined to be the spatial variation of the sky intensity in the limit of infinite integration time, and is found to be consistent among the different fields in our survey at the level of 5.8, 6.3 and $6.8\,$mJy/beam at 250, 350 and $500\,\mu$m, respectively.  These results, together with the measured instrument noise, may be used to estimate the integration time required for confusion limited maps, and provide a noise estimate for maps obtained by SPIRE.} 


  \maketitle
%

\section{Introduction}

The Spectral and Photometric Imaging Receiver (SPIRE)  (\cite{griffin}) onboard the 
{\it Herschel} Space Observatory (\cite{pilbrat}) has opened a new window on the Universe at far-infrared (FIR) wavelengths.  The sensitivity of SPIRE detectors combined with {\it Herschel}'s 3.5 meter aperture allow astronomers to observe the FIR sky with unprecedented
efficiency.   With 18.1", 24.9", 36.6" (FWHM) beams at 250, 350 and 500 microns, respectively, we expect SPIRE maps to be dominated by confused sources.   It is therefore useful to determine the key characteristics, both
of the instrument and the sky, that would allow observers to optimize their observing
plans, and/or to make sky surveys that probe as deep and as wide as possible
for allocated observing time.  In this letter we report measurements in SPIRE maps of the following: i.) instrument noise, i.e.
noise from the detectors, readout electronics and photon noise from the telescope, ii.) confusion noise, that is, the
variance in the sky map  due to the presence of unresolved sources, and iii.)  cosmic variance, arising from
underlying large-scale fluctuations in the galaxy number density.  

The confusion noise due to sources below a given flux cutoff, $x_c$, is derived in \cite{condon} to be the second moment of the measured flux distribution:
\begin{equation}
\centering
\sigma_{conf}^2 = \int_0^{x_c} x^2 dn \, ,
\end{equation}
where x is the measured flux, $x = Sf(\theta,\phi)$, $S$ is the source flux convolved with the normalized beam response, $f(\theta,\phi)$, and $dn$ is the differential source distribution.  In this analysis, the measured fluxes have been binned into sky map pixels and $\sigma_{conf}$  corresponds to  the variance in a sky map in the limit of
zero instrument noise, or equivalently, infinite integration time.  This physical
definition allows us to derive a direct, simultaneous measurement of both the confusion noise, $\sigma_{conf}$, and instrument noise, $\sigma_{inst}$, which then allow us to define the SPIRE confusion limit.  

In Sect. 2, we describe the observations, followed by the analysis details and results in Sect. 3.  In Sect. 4 we provide the number of ``repeats'' required to achieve confusion limited maps.   A repeat, as defined by the ``{\it Herschel} Space Observatory Observation and Proposal Submission Tool"  or HSPOT, contains two orthogonal scans of the entire field.  At nominal scan speed, a single repeat averages to 16, 29 and 28 samples/pixel, or an integration time of 0.9, 1.6 and 1.5 seconds/pixel.  In addition, we discuss the often-used (albeit, model-dependent) definition of confusion limit in terms of the source density or number of sources per 30 to 40 beams. 

\section{Data Sets}

The {\it Herschel} Multi-tiered Extragalactic Survey (HerMES\footnote{hermes.sussex.ac.uk}) Science Demonstration Phase (SDP) observations are detailed in Table 1 of \cite{Oliver}.  In this letter we have measured the noise properties of SPIRE maps of three of the five fields, GOODS-N,  Lockman-North and Lockman-SWIRE.  The GOODS-N field is 30'x30' and was covered by 30 map repeats.  The Lockman-North field is 35'x35' covered by 7 repeats.  These fields were observed in nominal scan mode with the spacecraft scanning at 30''/s with an angle of $\pm$ 42.4 degrees with respect to the spacecraft y-axis (see SPIRE Observers Manual (2010)).  Lockman-SWIRE is 218'x218' wide and was observed in SPIRE fast scan with a 60''/s scan speed, and was covered by 2 repeats.    Standard SPIRE pipeline maps (see Oliver et al., in prep.) were used.  These maps are calibrated in mJy/beam and  have pixel dimensions of 6", 10" and 14" at 250, 350 and 500 $\mu$m, respectively.  
   \begin{figure*}
   \centering
  \resizebox{\hsize}{!}{\includegraphics{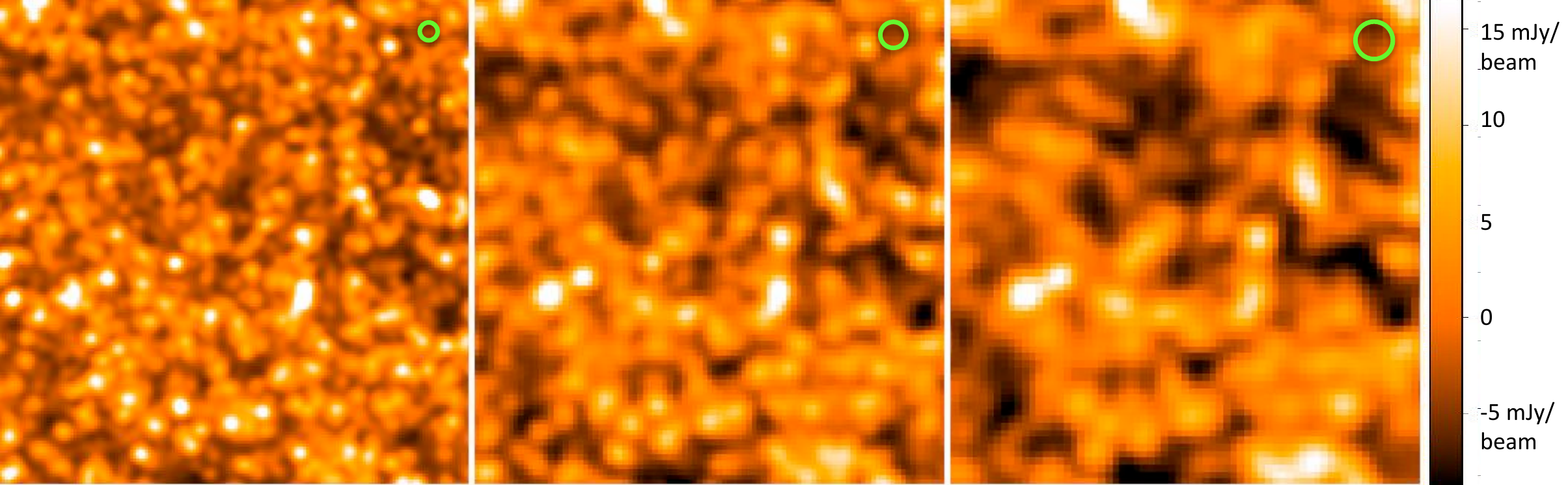}}
\caption{The GOODS-N field at 250, 350 and 500 $\mu$m (left to right).  Only the central 16'x16' of the 30'x30' field is shown above, with the green circles indicating the beam FWHM in each band.  As expected, many sources can be seen at 250 $\mu$m that are not resolved at longer wavelengths due to the larger beams as many of the 250 $\mu$m sources would be resolved with a smaller beam.  These images are a good illustration of confusion.}
         \label{goods_image}
   \end{figure*}
The two remaining HerMES SDP fields, FLS and Abell 2218 were not used in this analysis.  The FLS was contaminated by Galactic cirrus which is clearly seen in the measured pixel variance.
Abell 2218 is a small cluster field with lensing effects.  Both of these fields are strongly biased to higher confusion noise.

The central 16'x16' of the GOODS-N field are shown in Fig.~\ref{goods_image} at 250, 350 and 500 $\mu$m, clearly demonstrates the effect of confusion in SPIRE maps.  We note that the confusion noise according to eq. (1) is defined with some cutoff flux, $x_c$, which is set by the brightest sources in these maps.   We will discuss the effect of these cutoffs to the determination of the confusion noise.

\section{Analysis}
\label{analysis}

Fundamentally, spatial fluctuations in a map arise due to two sources: instrument noise and the presence of sources on the sky.  The instrument component, $\sigma_{inst}$, will be reduced with integration time, while fluctuations due to the convolution of the sky with the instrument beam, $\sigma_{conf}$,  will remain.  Since the sources of noise are uncorrelated, and if low-frequency correlated noise has been properly accounted for in the low level data reduction, the total variance in the map is simply, $\sigma_{total}^2$ = $\sigma_{conf}^2$ + $\sigma_{inst}^2/t$, where $t$ is the integration time.  Note that $\sigma_{conf}$ has unit of surface brightness in mJy/beam, while $\sigma_{inst}$ in mJy/beam$\sqrt{s}$.  A straight line fit of the variance vs. inverse integration time has a slope that determines the instrument variance, $\sigma_{inst}^2$, and a non-zero intercept that  determines the variance of the sky intensity, $\sigma_{conf}^2$ (\cite {jar03}).

\begin{figure*}
   \centering
   \includegraphics[width=9cm]{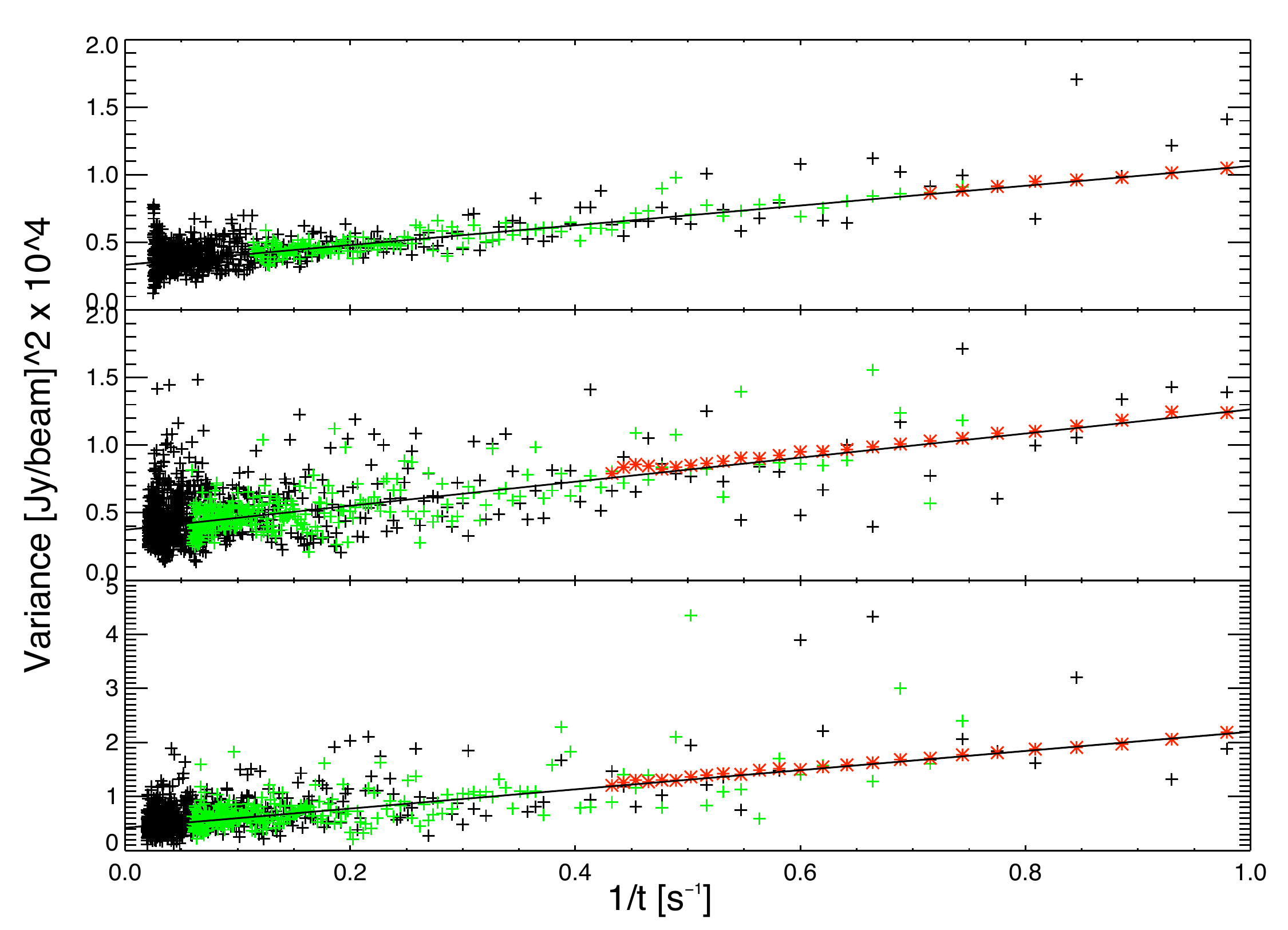}
   \includegraphics[width=9cm]{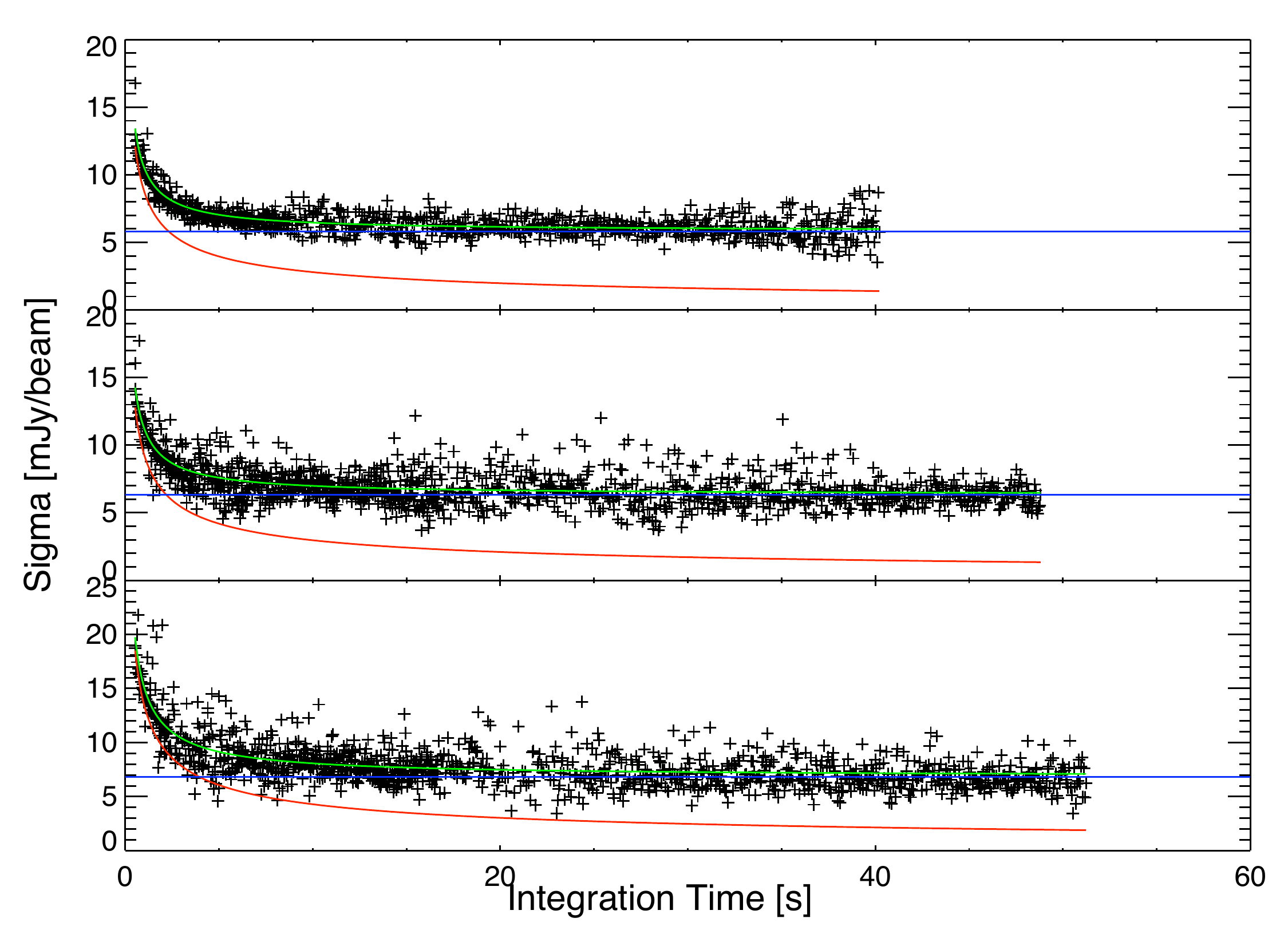}
      \caption{LEFT: Variance of SPIRE map pixels at 250, 350 and 500 $\mu$m (top to bottom) in 6", 10" and 14" pixels, respectively, in the GOODS-N (black), Lockman-North (green) and Lockman-SWIRE (red) fields.  Lockman-SWIRE was only observed once in each scan orientation, and was observed in Fast Scan Mode (60''/s). Despite the shallow field depth and different scan speed, the linear trend with inverse integration time is clearly continuous from short to long integration times.  The black line is a linear fit to all pixels in the three fields simultaneously revealing an instrument noise of 8.5  $\pm$ 0.4, 9.4  $\pm$ 0.5 and 13.3  $\pm$ 0.7 mJy/beam$\sqrt{s}$, and a confusion noise of 5.8 $\pm$ 0.3, 6.3  $\pm$ 0.4 and 6.8  $\pm$ 0.4 mJy/beam at 250, 350 and 500 $\mu$m. RIGHT: Pixel noise vs. integration time for all pixels in both fields. Overplotted are the derived instrument noise (red) and the confusion floor (blue) and the total noise (green).}
         \label{noise_all}
   \end{figure*}

In our analysis we begin by selecting all map pixels with a total integration time between $t - dt < t < t + dt$, where dt is 0.5 times the integration time per sample (i.e. 0.054 s/sample in nominal scan mode).   We then measure the variance, $\sigma_{total}^2$, of the pixel values in that subset.  Figure~\ref{noise_all} shows $\sigma_{total}^2(t)$ vs. $t^{-1}$ at 250, 350 and 500 $\mu$m  in the fields observed in nominal and fast scan mode. GOODS-N pixel sets are shown in black, Lockman-North in green and the shallow Lockman-SWIRE field in red.  The black line shows a simultaneous linear fit to all three fields and traces a component proportional to $t^{-1}$ and a non-zero intercept.   The data from the three included fields clearly show that the noise properties of the SPIRE instrument and the SPIRE observed sky are independent of the SPIRE scanning speed and stable from shallow to deep fields.   The consistency of the result in deep and shallow fields confirms that this method of noise characterization measures the confusion noise in maps that, themselves, are not confusion limited.  In fact, the large area shallow field (red points) in Fig.~\ref{noise_all} show far less scatter about the fit than the fewer pixels in the deep fields since there are many more pixels, providing a larger statistical ensemble.  

The right side of Fig.~\ref{noise_all} shows the same data in a different graphical representation.  The square root of the variance in all three fields (now all in black) is plotted vs. the integration time along with the derived instrument noise (red line) and confusion noise (blue line).  If the sky were dark, the noise would integrate down to zero following the red line, while for the real sky the noise can not get below the confusion floor.  

   \begin{table}
\begin{minipage}[t]{\columnwidth}
      \caption[]{Measured SPIRE Noise}
         \label{results}
\centering                                      
\renewcommand{\footnoterule}{}  
\begin{tabular}{lcccc}          
\hline\hline                        
Band & $\sigma_{conf}$ & $\sigma_{inst}$ &   \multicolumn{2}{c}{$\sigma_{inst}$} \\    
 $\mu$m & mJy/beam & mJy/beam$\sqrt{s}$ &  \multicolumn{2}{c}{mJy/beam$\sqrt{repeats}$} \\    
 & &  &  Nominal\footnote{30"/s scan speed} & Fast\footnote{60"/s scan speed}  \\    
\hline                                   
 250     &   5.8   $\pm$ 0.3   & 8.5   $\pm$ 0.4 & 9.0 &    12.7     \\
 350     &   6.3   $\pm$ 0.4   & 9.4    $\pm$ 0.5 & 7.5 &      10.6   \\
 500     &   6.8   $\pm$ 0.4   & 13.3   $\pm$ 0.7  & 10.8 &    15.3   \\
\hline                                             
\end{tabular}
\end{minipage}
\end{table}
The measured noise is presented in Table 1.  The confusion noise values in Table 1 are equivalent to the $1\sigma$ point source sensitivities, in mJy, in confusion limited maps.  The two rightmost  columns are included for observation planning and show the instrument noise scaled from mJy/beam$ \sqrt{s}$ to mJy/beam$ \sqrt{repeats}$ for nominal and fast scan mode, respectively, by averaging the total integration time per pixel in a map for a given number of map repeats.  Fast scan mode is $\sqrt{2}$ higher since it has half the samples in a sky pixel for a given number of map repeats. 

We have made 10 simulated realizations of our survey fields by 
constructing timestreams with known instrument noise and injecting 
sources drawn from the number counts determined by BLAST (Patachon
et al. 2009).  These simulations are used to  check for  biases due to the non-
gaussian nature of the pixel distribution or correlated noise and to 
provide better estimated uncertainties in our measurement.
These simulated timestreams (Glenn et al., in prep.)
are then processed through the mapping pipeline, and
the above variance analysis is run on the simulated maps. The
confusion and instrument noise measured in the simulated maps
agree with the analytically computed input noise in all three
bands, indicating that any bias is insignificant. The consistency of our 
results in the  deep and shallow fields further indicates an absence of 
measurement bias due to non-gaussianity.  The scatter in the instrument
and confusion noise determined in these simulations about the
known inputs indicates that the statistical uncertainty on the instrument
noise is  5\% and on the confusion noise  6\%, with
no significant bias in the recovered values. 

The agreement among the three fields of varying depth and solid angle also indicate that an additional uncertainty due to cosmic variance is unnecessary.  Cosmic variance, often negligible in relatively large surveys, can become a significant source of uncertainties in deep pencil-beam surveys for high redshift studies.  As a check, we re-ran the variance analysis on sub-fields of the GOODS-N field to estimate the field size at which the noise estimates diverge from those made from the full dataset.  That analysis shows that the confusion noise reported here is unaffected by cosmic variance in fields larger than  8'x8', 10'x10' and 13'x13' at 250, 350 and 500 $\mu$m.  As our smallest map is 30'x30', we include no uncertainty due to cosmic variance.

   \begin{figure}
   \centering
  \resizebox{\hsize}{!}{\includegraphics{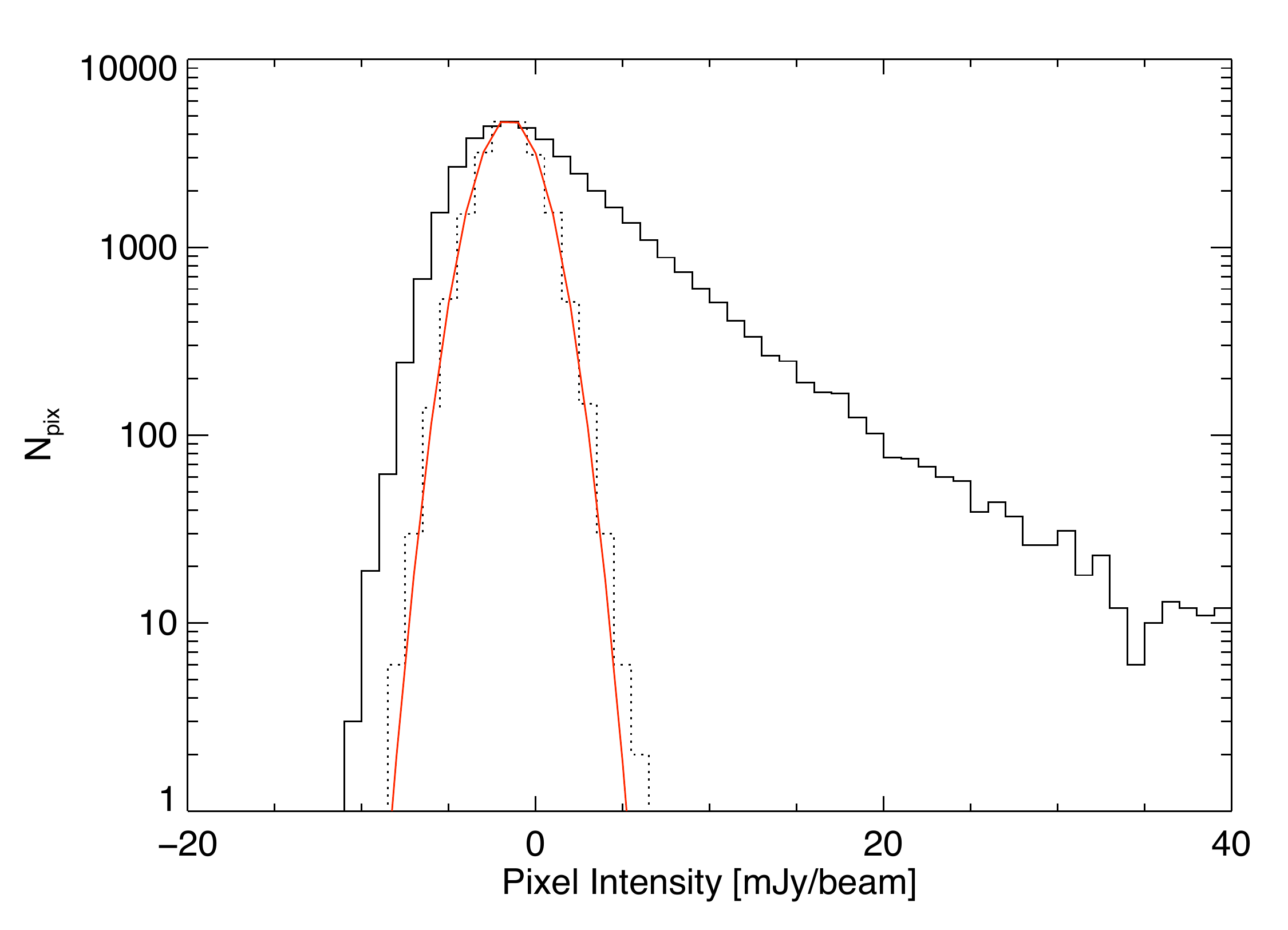}}
      \caption{Histogram of the 250 $\mu$m GOODS-N map, in black. The null map or jackknife removes the sky contribution and gives a map of the instrument noise, shown in the dashed histogram with the gaussian fit in red.  The instrument noise is integrated down to 1.7 mJy/beam, or 9.3 mJy/beam $\sqrt{repeats}$.  Similar analyses at 350 and 500 $\mu$m give instrument noise of 1.4 and 1.9 mJy/beam or 7.7 and 10.4 mJy/beam $\sqrt{repeats}$, consistent with the values in Table 1.}
         \label{hists}
   \end{figure}
Still, the accuracy of these results may depend on the data product or the quality of the map.   For independent verification, further analyses have been performed using an independent map maker (SMAP iterative map maker, Levenson et al., in prep.). Two maps of the GOODS-N field were made, one using the first 15 repeats and one with the second 15 repeats. A null map or jackknife was then created by subtracting these two maps.   Figure~\ref{hists} shows the pixel histogram of the GOODS-N map along with the first/second half jackknife map at 250 $\mu$m.  The jackknife should remove sky signal and leave only instrument noise.  Fitting a gaussian to the noise map histograms for GOODS-N maps with various numbers of repeats yields an instrument noise of 9.0 $\pm$ 0.1, 7.2 $\pm$ 0.1 and 10.4 $\pm$ 0.4 mJy/beam$\sqrt{repeats}$ at 250, 350 and 500 $\mu$m.  These values are consistent with those measured using the variance method described above (see Table 1) and, as the variance method gives a precision measurement of the confusion noise as well, we present the results of that analysis as the fiducial SPIRE noise estimates.  

\begin{figure}
   \centering
  \resizebox{\hsize}{!}{\includegraphics{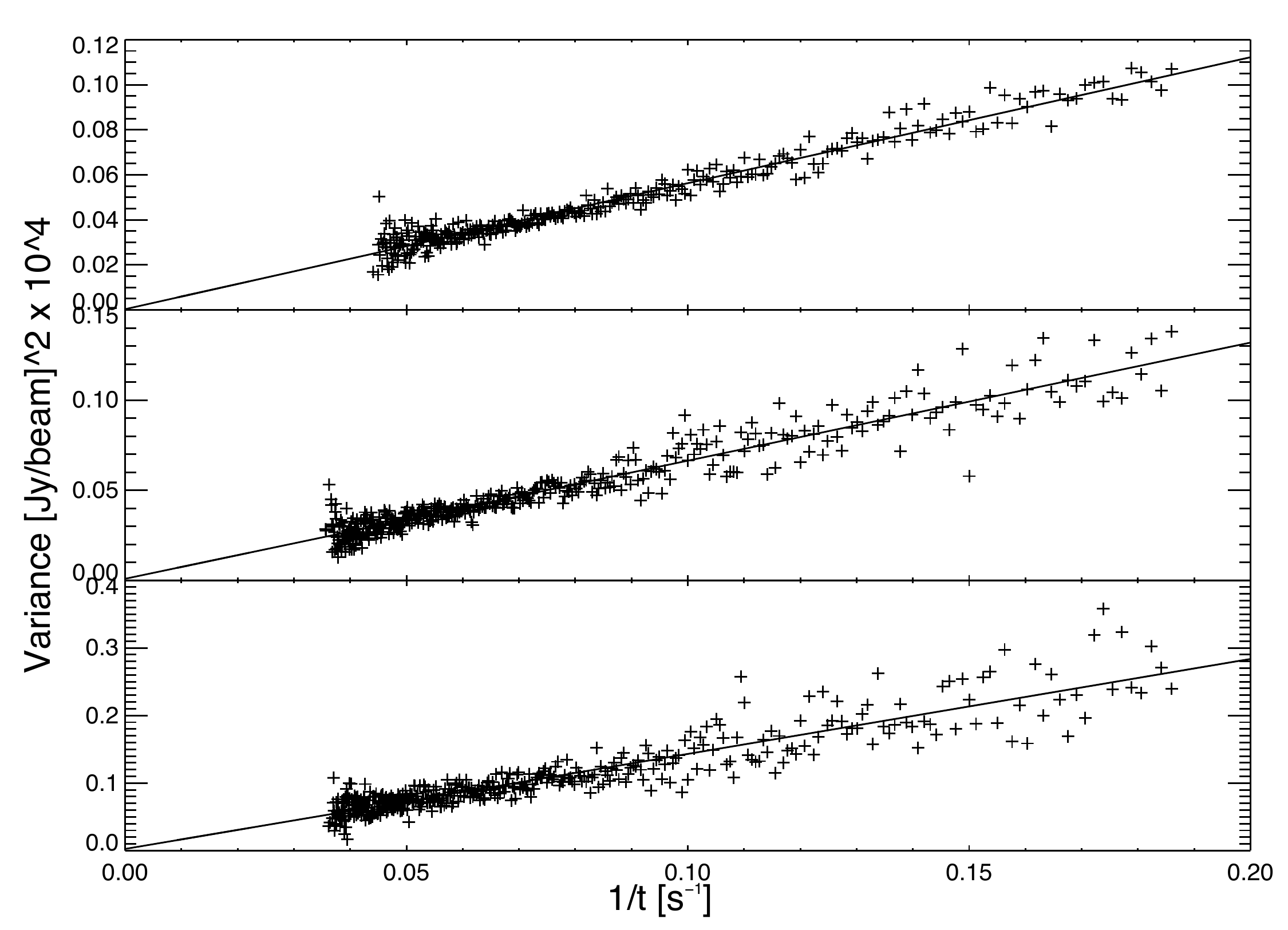}}
      \caption{Variance plot for jackknife map of GOODS-N.  As expected, the null map gives a linear fit with the similar slope or instrument noise as the full map, and the intercept or confusion noise is consistent with zero since sky contribution has been removed.}
         \label{noise_jack}
   \end{figure}

Figure~\ref{noise_jack} shows the equivalent of left column of Fig.~\ref{noise_all} for the jackknife map of GOODS-N.  As expected, the instrument noise remains in the jackknife map, and with the sources removed the intercept is now consistent with zero.  Specifically, intercepts from jackknife map give noise estimates of 0.3 $\pm$ 0.3, -0.3 $\pm$ 0.3 and -0.4 mJy/beam $\pm$ 0.5 at  250, 350 and 500 $\mu$m.  Instrument noise measured using the same method in the jackknife map is 8.4 $\pm$ 0.9, 9.0 $\pm$ 1.1 and 12.8 $\pm$ 1.8 mJy/beam$\sqrt{s}$ which agrees with the values determined from the full maps.  

\section{Discussion}

We have made a precision measurement of the confusion and instrument noise in the HerMES SDP SPIRE scan maps.  The various checks presented, including simulated datasets with known noise properties and a repeat of the analysis on null map or jackknife, demonstrate the robustness of the analysis and give good consistency in the values of the confusion and  instrument noise.  Our simulations indicate that the statistical error of the confusion noise is about 5$\%$.  We note that the systematic error is largely dependent on the calibration and map-making process, currently estimated to be $\sim$ 15$\%$ (\cite{griffin}).  Table 1 can be used as a guide for achieving confusion limited maps with SPIRE.  In particular, to make maps in which the instrument noise is comparable to the confusion, it will take a minimum of 3 map repeats in nominal scan mode or 5 map repeats in fast scan mode.  Longer integration time will result in maps in which confusion is the dominant source of uncertainty in measurements of source flux and position.

\begin{figure}
   \centering
  \resizebox{\hsize}{!}{\includegraphics{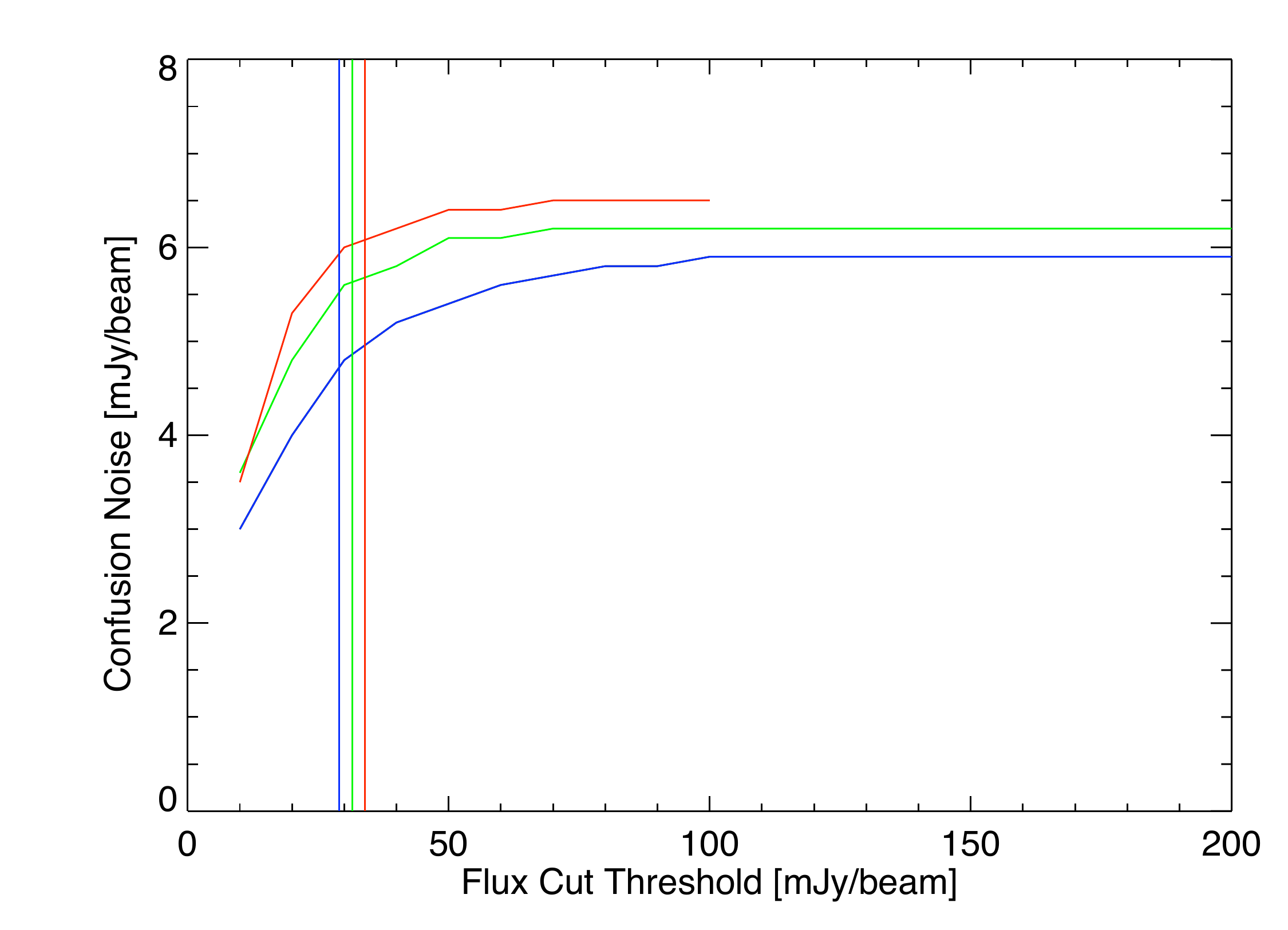}}
      \caption{Confusion noise as a function of pixel flux cut at 250/350/500 $\mu$m in blue/green/red.  All pixels within a beam FWHM of a pixel brighter than the indicated flux cut were masked and the confusion noise was re-estimated.  Vertical lines indicate 5$\sigma_{conf}$ in each band.}
         \label{cuts}
   \end{figure}

The noise values reported in Table 1 make no significant source cut and, accounting for the possibility that even bright sources are confused, measure the variance in HerMES maps due to all sources up to 10$\sigma_{conf}$ or 80 mJy (more than 99\% of data).  It may be interesting to systematically remove bright sources and study how confusion noise might change accordingly.  We have determined the confusion noise using the same method presented in Sect. 3, after removing pixels within a beam FWHM of any pixel brighter than a given flux cut.  (Since the maps are calibrated such that the source flux is given by the flux in the pixel at the peak of the PSF, i.e., mJy/beam, this method is equivalent to removal of sources.  Removing pixels within a FWHM is conservative and ensures that any
extended emission is completely removed.) The results of this analysis are shown in Fig.~\ref{cuts}.  The vertical lines indicate 5$\sigma_{conf}$ as measured in the full map.  The residual noise estimates are 3.8, 4.6 and 5.2 mJy with a 3$\sigma_{conf}$ cut and 4.8, 5.5 and 6.1 mJy with a 5$\sigma_{conf}$ cut.  

The confusion limit is often quoted in terms of the flux, at which the source density in a map
reaches 30 to 50 beams per source. This quantity requires a measurement or a model of the density
of sources, and often is derived assuming a power law source distribution with a 3 -4$\sigma_{conf}$ (\cite{frances}).  BLAST (Devlin et al. 2009) predicts confusion limits corresponding to 40 beams per source for SPIRE of 22, 22 and 18 mJy at 250, 350 and 500  $\mu$m, respectively.
The best source counts available at the SPIRE wavelengths
are probably the HerMES counts themselves, as reported
by Oliver et al. (2010). The HerMES counts reach 1 source per 40
beams at $19.1\pm0.6$, $17.7\pm0.6$, and $15.1\pm 1.8$
mJy at 250, 350 and 500 $\mu$m,  corresponding to 3.29, 2.81, and 
2.60$\sigma_{conf}$, and in rough agreement with BLAST's prediction.

\begin{acknowledgements}
SPIRE has been developed by a consortium of institutes led by
Cardiff Univ.\ (UK) and including Univ.\ Lethbridge (Canada);
NAOC (China); CEA, LAM (France); IFSI, Univ.\ Padua (Italy);
IAC (Spain); Stockholm Observatory (Sweden); Imperial College
London, RAL, UCL-MSSL, UKATC, Univ.\ Sussex (UK); Caltech, JPL,
NHSC, Univ.\ Colorado (USA). This development has been supported
by national funding agencies: CSA (Canada); NAOC (China); CEA,
CNES, CNRS (France); ASI (Italy); MCINN (Spain); SNSB (Sweden);
STFC (UK); and NASA (USA). 



\end{acknowledgements}

\end{document}